\documentclass[preprint,review,1p]{elsarticle}
 \usepackage{amssymb}
 \usepackage{amssymb}
 \usepackage{amsmath}
 \usepackage{graphicx,color}
 \usepackage{epstopdf}
 \usepackage{caption}
\usepackage{subcaption}

\journal{International Journal of Non-Linear Mechanics}

\begin{document}

\begin{frontmatter}

 \title{Traveling waves in one-dimensional nonlinear models of  strain-limiting  viscoelasticity}

\author{H. A. Erbay}
    \ead{husnuata.erbay@ozyegin.edu.tr}
\author{Y. \c Seng\"{u}l\corref{cor1}}
    \ead{yasemin.sengul@ozyegin.edu.tr}

\cortext[cor1]{Corresponding author. Tel: +90 216 564 9498 Fax: +90 216 564 9057}

\address{Department of Natural and Mathematical Sciences, Faculty of Engineering, Ozyegin University,  Cekmekoy 34794, Istanbul, Turkey}

  \begin{abstract}
 In this article we investigate traveling wave solutions of a nonlinear differential equation describing the behaviour of one-dimensional viscoelastic medium with implicit constitutive relations. We focus on a subclass of such models known as the strain-limiting models introduced by Rajagopal.  To describe the response of viscoelastic solids we assume a nonlinear relationship among the linearized strain, the strain rate and the Cauchy stress. We then concentrate on traveling wave solutions that correspond to the heteroclinic connections between the two constant states. We establish conditions for the existence of such solutions, and find those solutions, explicitly, implicitly or numerically, for various forms of the nonlinear constitutive relation.
 \end{abstract}


\end{frontmatter}

\setcounter{equation}{0}
\section{Introduction}
\noindent

The present paper is  concerned with the dynamics of a viscoelastic medium investigating the traveling wave solutions of the equation
\begin{equation}\label{model}
    T_{xx} + \nu\, T_{xxt} = g(T)_{tt},
\end{equation}
where $T(x, t)$ is the Cauchy stress at point $x$ and time $t$, $g(\cdot)$ is a nonlinear function and $\nu > 0 $ is a constant.
Equation \eqref{model} is a one-dimensional nonlinear differential equation in $T$ resulting from the equation of motion and a constitutive equation relating the stress, the linearized strain and the strain rate.

As opposed to the classical models in mechanics, the strain can be written as a function of the stress, rather than expressing the stress in terms of the kinematical variables. This idea is due to Rajagopal \cite{Rajagopal-2003,Rajagopal-2007}, who introduced a generalization of the theory of elastic materials by suggesting implicit models allowing for approximations where the linearized strain is a nonlinear function of the stress. A series of papers on such implicit theories has been published recently (see e.g. \cite{BuMaRaSu}, \cite{Bus-2009}, \cite{Bus-Raj-2011}, \cite{Rajagopal-2014}, \cite{Raj-Sac-2014}). The advantage of this  new idea is that it allows for the gradient of the displacement to stay small so that one could treat the linearized strain, even for arbitrary large values of the stress. In this work we focus on four different such models, and we reconsider them in the context of viscoelasticity. We also look at models with quadratically and cubically nonlinear constitutive relations although they do not behave as expected for large values of the stress.

There are numerous models introduced by Rajagopal in \cite{Rajagopal-2003} with implicit constitutive relations between the stress and the strain including models for elastic fluids, inelastic materials and non-hyperelastic materials. Following these models, various forms of nonlinear constitutive relations have been studied in different contexts. For example, Kannan, Rajagopal and Saccomandi \cite{Kannan-Raj-Sac} worked on the elastic case with a polynomial type nonlinearity (see Section \ref{sec:1D-visc} for more details). Bul\'{\i}\v{c}ek et al. \cite{BuMaRaSu}, on the other hand, considered the static case with a more general nonlinearity (see Section \ref{sec:1D-visc}) and presented the first existence result in a three-dimensional domain.

For viscoelasticity, much less is done in the literature. As explained by Muliana, Rajagopal and Wineman in \cite{Mul-Raj-Wine}, force, and hence the stress, is the cause for deformation, hence for the strain. Because of this the strain should be described in terms of the stress or its history than vice versa. The motivation for this idea is that in the classical elasticity theory, there cannot be a nonlinear relationship between the linearized strain and the stress, which, in fact, is observed in some experiments (see e.g. \cite{Saito-et-all}, \cite{Rajagopal-2014}). The fracture of brittle elastic bodies is another possible application area for such implicit theories, where one can obtain bounded strain at the crack tip due to the possibility of having a nonlinear relationship between the linearized strain and the stress (see \cite{Rajagopal-Walton} for details). Muliana et al. \cite{Mul-Raj-Wine} developed a quasi-linear viscoelastic model where the strain is expressed as an integral of a nonlinear measure of the stress. Rajagopal and Srinivasa in \cite{Rajagopal-Srinivasa} proposed a Gibbs-potential-based formulation for the response of viscoelastic materials in this new class. Also Rajagopal and Saccomandi \cite{Raj-Sac-2014} investigated viscoelastic response of solids, a one-dimensional version of which is the one we study in this work, namely
\begin{equation}\label{Raj-Sac-general}
\gamma \mathbf{B} + \nu \mathbf{D} = \beta_{0} \mathbf{I} + \beta_{1} \mathbf{T} + \beta_{2} \mathbf{T}^{2},
\end{equation}
where $\gamma$ and $\nu$ are nonnegative constants, $\beta_{i} = \beta_{i}(I_{1}, I_{2}, I_{3}), (i = 0, 1, 2)$, $I_{1}  = \mathrm{tr} \mathbf{T}, I_{2} = \frac{1}{2} \mathrm{tr} \mathbf{T}^{2}, I_{3} = \frac{1}{3} \mathrm{tr} \mathbf{T}^{3}$, $\mathbf{B}$ is the left Cauchy-Green stretch tensor and $\mathbf{D}$ is the symmetric part of the gradient of the velocity field. As they explain, this model includes as special subcases; models for a very general new class of elastic and viscoelastic bodies (e.g. Titanium and Gum metal alloys), as well as the Navier-Stokes fluid model (see e.g. \cite{Rajagopal-2007}). Linearizing the strain in this model reduces \eqref{Raj-Sac-general} to
\begin{equation}\label{Raj-Sac-model}
\boldsymbol{\epsilon} + \nu \boldsymbol{\epsilon}_{t} = \beta_{0} \mathbf{I} + \beta_{1} \mathbf{T} + \beta_{2} \mathbf{T}^{2},
\end{equation}
where $\boldsymbol{\epsilon} = \frac{1}{2} (\mathbf{\nabla} u + \mathbf{\nabla} u^{T})$ is the linearized strain, and $u(x,t)$ is the displacement.

We study \eqref{Raj-Sac-model} in one space dimension with a general nonlinear right-hand side (see \eqref{cons-eqn}). We are interested in analyzing the conditions on the nonlinearity $g(T)$ when traveling wave solutions of the form $T(\xi)$ with $\xi = x - c t,$ where $c$ represents the wave propagation speed, exist for two constant equilibrium states at infinity. We find the solutions analytically (implicitly or explicitly) or numerically. More precisely, we will first look at the quadratic and the cubic cases for which we are able to solve the problem analytically and obtain explicit or implicit solutions. After that we will study four nonlinear models, namely Models A, B, C and D (see Section \ref{sec:1D-visc}), and we will either express the solution implicitly, or obtain it numerically if it is not possible to find an analytical solution. Our work seems to be the first such treatment in the literature of strain-limiting viscoelasticity.

The propagation  of traveling waves in nonlinear viscoelastic solids has also been  studied previously in the context of classical theory of viscoelasticity (see e.g. \cite{Destrade-Saccomandi}, \cite{Destrade-Jordan-Saccomandi}, \cite{Jordan-Puri}, and references therein).  The results of present work  exhibit some similarities with those in the literature. The first common point is  that  the equations of motion admit  kink-type traveling wave solutions. Also, in both cases, the effective width of the traveling wave  is proportional to the viscosity parameter and  the wave profile becomes smoother as the viscosity parameter increases. However, our study differs from the articles  within the context of classical viscoelasticity theory in the sense that the governing equation in our model (see (1.1)) is in terms of the stress and also the nonlinearity is on the inertia term.

The structure of  the paper is as follows. In Section \ref{sec:1D-visc} we introduce the one-dimensional strain-limiting viscoelasticity model as well as give a list of four nonlinear constitutive relations that has been suggested for elastic solids. In Section \ref{sec:traveling-wave} we consider traveling wave solutions of the governing equations. In Section \ref{sec:applications} we solve the resulting differential equation for different nonlinear constitutive relations, and give analytical solutions where possible, or obtain numerical solutions.

\setcounter{equation}{0}
\section{One-dimensional strain-limiting viscoelasticity}\label{sec:1D-visc}
\noindent

Consider a one-dimensional, homogeneous, viscoelastic, infinite medium exhibiting small strains for large stresses.   In the absence of external body forces, the equation of motion is given by
\begin{equation}
    \rho_{0} u_{tt} = T_{x},   \label{eqn-motion}
\end{equation}
where  $\rho_{0}$ is the mass density of the medium, the scalar-valued function $u(x,t)$ is the displacement, and  $T(x,t)$ is the Cauchy stress. Here and throughout this work the subscripts denote partial derivatives. In contrast to explicit constitutive relations of the classical theories of viscoelasticity, we shall employ an implicit constitutive relation
\begin{equation}
    \epsilon + \nu \epsilon_t = g(T),   \label{cons-eqn}
\end{equation}
which gives the linearized strain  $\epsilon = u_{x}$ and the strain rate  $\epsilon_t$ as a nonlinear function of the stress $T$, with  $g(0) = 0$ and a nonnegative constant $\nu$. The model defined by \eqref{cons-eqn} is the one dimensional form of \eqref{Raj-Sac-model}. When $\nu=0$, it reduces to the one-dimensional version of the model  introduced by Rajagopal in \cite{Rajagopal-2003, Rajagopal-2007} for elastic solids.

For convenience, we now define the dimensionless quantities
\begin{equation}
    \bar{x} = \frac{x}{L}, ~~~~\bar{t} = \frac{t}{L} \sqrt{\frac{\mu}{\rho}}, ~~~~\bar{T} = \frac{T}{\mu}, ~~~~
    \bar{u} = \frac{u}{L}, ~~~~ \bar{\nu} = \frac{\nu}{L} \sqrt{\frac{\mu}{\rho}}, \label{dimen}
\end{equation}
where $L$ is a characteristic length and $\mu$  is a constant with the dimension of stress. Differentiating both sides of \eqref{eqn-motion} with respect to $x$,  substituting  \eqref{cons-eqn} into the resulting equation and using \eqref{dimen}, we obtain \eqref{model}, where we drop the overbar for notational convenience. The question that we shall discuss throughout the rest of this work is which of the possible forms of the nonlinear function $g(T)$ are relevant for the existence of traveling wave solutions of \eqref{model}. Following mainly the standard techniques used widely in the literature to find traveling wave solutions we obtain the solutions of \eqref{model}, explicitly, implicitly or numerically, for various forms of $g(T)$.

We now discuss some strain-limiting models  reported in the literature for  elastic and viscoelastic solids. The following is a list of nonlinear constitutive relations $g(T)$ which we adopt in this study.

\textbf{Model A} : We first consider the one-dimensional version of the model proposed in an elastic setting by Kannan, Rajagopal and Saccomandi in \cite{Kannan-Raj-Sac}, namely,
\begin{equation}\label{modelA}
g(T) = \beta T + \alpha \left(1 + \frac{\gamma}{2} T^{2}\right)^{n} T,
\end{equation}
where $\alpha \geq 0$, $\beta \leq 0$, $\gamma \geq 0$ and $n$ are constants. Note that when $n=0$ and/or  $\gamma = 0$, one recovers the standard constitutive equation for a linearized material.  In Section \ref{sec:applications}, for the strain-limiting viscoelastic  model defined by \eqref{cons-eqn}-\eqref{modelA}    we obtain traveling wave solutions  explicitly if $n = 1$ and implicitly if $n = -1/2$.

\textbf{Model B} :  The second model is based on a simplified version of the nonlinear constitutive relation proposed by Rajagopal in \cite{Rajagopal-2011}:
\begin{equation}\label{modelB}
g(T) = \frac{T}{(1 + | T |^{r})^{1/r}},
\end{equation}
where $r>0$ is a constant. This model was studied in elastic settings by many authors in different contexts (see e.g. \cite{BuMaRaSu, BuMaRaWal, BuMaSu}). Note that when $\beta = 0$, $n=-1/2$, $\alpha=1$ and $\gamma=2$, Model A becomes equivalent to Model B with $r=2$. In Section \ref{sec:applications}, when $r=2$, traveling wave solutions corresponding to this model are obtained in closed form.

\textbf{Model C} :  This model is the one-dimensional form of the constitutive relation proposed by Rajagopal in \cite{Rajagopal-2010,Rajagopal-2011};
\begin{equation}\label{modelC}
    g(T) = \alpha \left\{\left[1 - \exp\left(- \frac{\beta T}{1 + \delta |T|}\right)\right] + \frac{\gamma T}{1 + |T|}  \right\},
\end{equation}
where $\alpha$, $\beta$, $\gamma$ and $\delta$ are constants. Note that when $\beta = 0$ and $\alpha=\gamma=1$ this model reduces to Model B with $r=1$. In Section \ref{sec:applications}, we solve the nonlinear differential equation corresponding to this model numerically and  compute traveling wave solutions for a specific set of parameter values.

\textbf{Model D} : This model is the one-dimensional form of a different model again introduced by Rajagopal in \cite{Rajagopal-2010,Rajagopal-2011};
\begin{equation}\label{modelD}
    g(T) = \alpha \left(1-\frac{1}{1 +\frac{ T}{1 + \delta |T|}}\right) + \beta \left(1 + \frac{1}{1 + \gamma T^{2}}\right)^{n} T,
\end{equation}
where $\alpha$, $\beta$, $\gamma$ and $\delta$ are constants. Note that when $\alpha = 0$, with appropriate choice of the remaining parameters, we may derive Model A from this model. In Section \ref{sec:applications}, traveling wave solutions corresponding to this model are also obtained numerically for a specific set of parameter values.

Before going further, we would like to recall the remark made by Rajagopal in \cite{Rajagopal-2010} about Models C and D. He says that both Model C and Model D have a drawback when the stress is compressive and sufficiently large. It is obvious from \eqref{modelC} and \eqref{modelD} that the assumption of small  strain will be violated due to the initial terms in these equations  when the stress is negative and sufficiently large. Furthermore, as it was also mentioned by Rajagopal in the same article,  there are typographical errors in equations (3.12) and (3.13) of \cite{Rajagopal-2011}, where Models C and D were introduced.

\begin{figure}[ht]
\begin{center}
    \includegraphics[width=200pt]{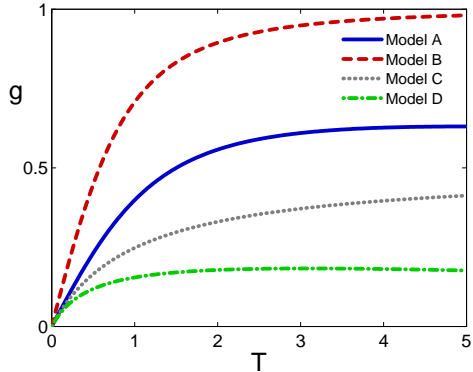}%
    \caption{Variation of the function $g(T)$ with moderate $T$ values for various nonlinear models of strain-limiting viscoelasticity. The specific sets of the parameter values are as follows: $\alpha=0.5$, $\beta=-0.01$, $\gamma=1$ and $n=-0.5$ for Model A, $r=2$ for Model B,  $\alpha=0.5$, $\beta=-0.01$ and $\gamma=\delta=1$ for Model C and $\alpha=0.5$, $\beta=-0.01$, $\gamma=\delta=1$ and $n=0.5$ for Model D.}
    \label{g-vs-T}
\end{center}
\end{figure}

Figure \ref{g-vs-T} shows the variation of $g(T)$ with $T$  in a moderate stress regime for the above mentioned four nonlinear models with some specifically chosen parameter values. We note that, due to  \eqref{cons-eqn}, the vertical axis in Fig. \ref{g-vs-T}  measures the sum of the linearized strain and the strain rate. Moreover,  we observe that in the case of  moderate stress levels the linearized strain may remain finite  for the above models of strain-limiting viscoelasticity depending on the parameter values.

\setcounter{equation}{0}
\section{Traveling wave solutions}\label{sec:traveling-wave}
\noindent

In this section we investigate  traveling wave solutions  of \eqref{model}. Traveling waves are solutions of the form
\begin{equation}
    T = T(\xi),~~~~ \xi = x - c t, \label{travel}
\end{equation}
where  the wave propagation speed $c$ is a constant to be determined below. Substitution of \eqref{travel} into \eqref{model} reduces the third-order partial differential equation to the third order ordinary differential equation in the variable $\xi$ given by
\begin{equation}\label{ode}
    T^{\prime\prime} - \nu\,c\, T^{\prime\prime\prime} = c^2 [g(T)]^{\prime\prime},
\end{equation}
where the symbol $^\prime$ stands for differentiation. For the rest of this study, we focus on traveling wave solutions of \eqref{ode} that correspond to the heteroclinic connections between two constant states. Obviously,  $T(\xi)\equiv \mbox{constant}$ is a trivial solution of \eqref{ode}, so we assume that
\begin{equation}\label{bound}
    \lim_{\xi \rightarrow -\infty} T(\xi)=T_{\infty}^{-}, ~~~~ \lim_{\xi \rightarrow +\infty} T(\xi)=T_{\infty}^{+}
\end{equation}
with $T_{\infty}^{-}\neq T_{\infty}^{+}$, where $T_{\infty}^{-}$ and $T_{\infty}^{+}$ are constants to be specified later. Our main problem is to find restrictions on the nonlinear function $g(T)$,  which guarantees the existence of such a traveling wave solution, and is to discuss, from this point of view,  the constitutive functions suggested in the literature.

We now integrate  \eqref{ode} once  and then use the boundary conditions $T^{\prime}(\xi), T^{\prime\prime}(\xi)\rightarrow 0$ as $\xi \rightarrow \pm \infty$, to eliminate the arbitrary integration constant. A further integration of the resulting equation yields
\begin{equation}\label{theEQN}
    T - \nu\,c\,T^\prime = c^{2} g(T) + A,
\end{equation}
where $A$ is an arbitrary integration constant. Boundary conditions \eqref{bound} then give
\begin{equation}\label{A}
    A = \frac{1}{2} \{T_{\infty}^{-} + T_{\infty}^{+} - c^{2}  [g(T_{\infty}^{-} )+ g(T_{\infty}^{+})]\},
\end{equation}
and
\begin{equation}\label{c-2}
    c^{2} = \frac{T_{\infty}^{-} - T_{\infty}^{+}}{g(T_{\infty}^{-}) - g(T_{\infty}^{+})}.
\end{equation}
Thus the squared wave speed is obtained in terms of the two known states at infinity. Using \eqref{A} to eliminate $A$ in \eqref{theEQN} we get the differential equation
\begin{equation}\label{general-ODE}
    T^\prime = f(T)
\end{equation}
where
\begin{equation*}
    f(T) = \frac{1}{\nu\,c}\left\{\left(T - \frac{T_{\infty}^{-} + T_{\infty}^{+}}{2}\right)
            - c^{2} \left[g(T) - \frac{g(T_{\infty}^{-}) + g(T_{\infty}^{+})}{2}\right]\right\}.
\end{equation*}
Of course, two obvious equilibrium points of \eqref{general-ODE} are $T=T_{\infty}^{-}$ and $T=T_{\infty}^{+}$, that is,  $f(T_{\infty}^{-})= f(T_{\infty}^{+})=0$. Integrating \eqref{general-ODE} we get the implicit solution in the form
\begin{equation}\label{implicit}
    \xi - \xi_{0} = \int_{T_{0}}^{T} \frac{ds}{f(s)},
\end{equation}
where $\xi_{0}$ is a constant and $T(\xi_{0}) = T_{0}$.

We conclude this section  with a description of a prototype problem on which we will discuss the consequences of various forms of the nonlinear function $g(T)$ in the next section. Recall that a heteroclinic traveling wave propagates from one constant state to the other if $c^{2}>0$. Due to  \eqref{c-2} this implies that a traveling wave solution of \eqref{ode}-\eqref{bound}  exists  in one of the following two cases:
\begin{equation}\label{con1}
    \mbox{Case (i)}~~~~~  T_{\infty}^{-} > T_{\infty}^{+}  ~~~\mbox{and}~~~ g(T_{\infty}^{-}) > g(T_{\infty}^{+}),
\end{equation}
or
\begin{equation}\label{con2}
    \mbox{Case (ii)}~~~~~  T_{\infty}^{-} < T_{\infty}^{+}  ~~~\mbox{and}~~~ g(T_{\infty}^{-}) < g(T_{\infty}^{+}).
\end{equation}
For the remainder of this paper, without loss of generality, we restrict our attention to the first case for tractability reasons. Furthermore, we assume that the two constant equilibrium states are a normalized state of the stress and the zero reference state of the stress; that is, we take
\begin{equation}\label{ass-pirs}
    T_{\infty}^{-} = 1 \quad \text{and} \quad T_{\infty}^{+} = 0.
\end{equation}
One should recall that the stress is dimensionless. Also, even though we restrict our attention to the case \eqref{ass-pirs} we should be aware that the nonlinearity amplifies the values of $g(T)$ when $|T| > 1$ and reduces them when $|T| < 1$, and, depending on which range of $T$ we are working in, the traveling wave profile for $g(T)$ is affected correspondingly.

 We note that \eqref{con1}, \eqref{ass-pirs} and $g(0)=0$ imply $g(1)>0$.  This condition is automatically satisfied by Model B (recall that  $r>0$), but it imposes the following restrictions on the parameters of Models A, C and D; namely,
\begin{equation*}
    g(1) =\beta +  \alpha \left(1 +\frac{\gamma}{2}\right)^{n}>0,
\end{equation*}
\begin{equation*}
    g(1) = \alpha \left\{\left[1 - \exp\left(- \frac{\beta}{1 + \delta }\right)\right] + \frac{\gamma}{2}  \right\}>0,
\end{equation*}
and
\begin{equation*}
    g(1) = \frac{\alpha}{2 + \delta} + \beta \left(\frac{2+\gamma}{1 + \gamma}\right)^{n} >0,
\end{equation*}
respectively.
Plugging \eqref{ass-pirs} into \eqref{c-2} gives
\begin{equation}\label{cc}
    c^{2} = 1/g(1)
\end{equation}
where we have used $g(0)=0$. With the use of \eqref{ass-pirs}, the differential equation \eqref{general-ODE} becomes
\begin{equation}\label{sample}
    T^\prime =  \frac{1}{\nu\,c\,g(1)}\left[g(1)T-g(T)\right],
\end{equation}
which is studied for various forms of $g(T)$ in the next section. We also note that solutions of \eqref{general-ODE} and \eqref{sample} are translational invariant. That is,  if $T(\xi)$ is a solution of \eqref{general-ODE} or \eqref{sample}, then so is $T(\xi+p)$ for any fixed constant $p$. Consequently, noting that $T(0)$ can take any number in the range of values for $T$, we fix the traveling wave solution by assuming that
\begin{equation}\label{T_0}
    T(0)  = 1/2.
\end{equation}

Two equilibrium points of \eqref{sample} are clearly $T=1$ and $T=0$ (since $g(0)=0$). Equation \eqref{sample} may have additional equilibrium points depending on the form of $g(T)$. Assume that \eqref{sample} has an equilibrium point $T^{*}$ for which $g(1)T^{*}=g(T^{*})$. The linearization of  \eqref{sample} at this  point possesses one real eigenvalue
\begin{equation*}
    \lambda=  \frac{g(1)-g^{\prime}(T^{*})}{\nu\,c\,g(1)},
\end{equation*}
which shows that $T^{*}$ is an unstable equilibrium for $g(1)\neq g^{\prime}(T^{*})$, and a stable equilibrium for $g(1)= g^{\prime}(T^{*})$.

\setcounter{equation}{0}
\section{Applications to some nonlinear models }\label{sec:applications}
\noindent

 This section  discusses in detail both quadratic and cubic models of strain-limiting  viscoelastic solids and the nonlinear models presented in Section 2, within the context of Section \ref{sec:traveling-wave}.

 We first remark that there is no heteroclinic traveling wave solution  when we consider an elastic solid for which  $\nu = 0$. This can be easily seen from \eqref{general-ODE} or \eqref{sample} by neglecting the derivative term (i.e. the dissipation term). Then, the only solution of the resulting algebraic equation is a constant solution but the boundary conditions at infinity require two different constants, giving a contradiction.

 A similar conclusion is also valid for the linear viscoelastic model for which we have $g(T) = g^\prime(0) T$ with $g^\prime(0) \neq 0$. In such a case \eqref{general-ODE} (or \eqref{sample}) reduces to $T^\prime = 0$ which implies $T $ is a constant.  Following the same line of reasoning we find that there is no heteroclinic traveling wave solution for the strain-limiting linear viscoelastic model.

\begin{figure}[ht]
\begin{center}
    \includegraphics[width=200pt]{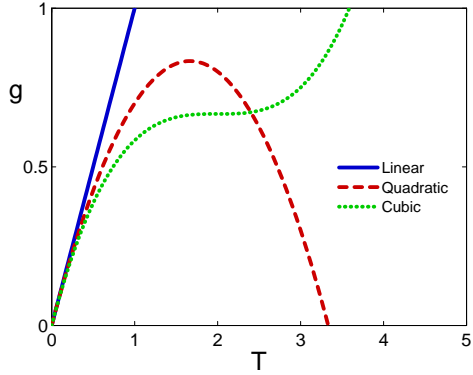}%
    \caption{Variation of the function $g(T)$ with moderate $T$ values for the linear, quadratic and cubic models of strain-limiting viscoelasticity. The specific sets of the parameter values are as follows: $g^{\prime}(0)=1$ for the linear model, $g^{\prime}(0)=1$ and $g^{\prime\prime}(0)=-0.6$ for the quadratic model, and $g^{\prime}(0)=1$, $g^{\prime\prime}(0)=-1$ and $g^{\prime\prime\prime}(0)=0.5$ for the cubic model.}
    \label{lin-quad-cub}
\end{center}
\end{figure}

In the remaining part of this section we focus on six particular forms of $g(T)$; the quadratic and cubic models, and  the nonlinear models described in Section \ref{sec:1D-visc}, namely Models A, B, C and D.  Figure \ref{lin-quad-cub} shows the variation of $g(T)$ with $T$ for linear, quadratic and cubic models in a moderate stress regime.   We observe from Fig. \ref{lin-quad-cub} that, depending on the chosen parameter values,  the  quadratic  and cubic models exhibit qualitatively different responses and they may give rise to negative or large positive values of $g(T)$  with increasing values of the stress. Obviously, the case where a positive (tensile) stress gives rise to a negative (compressive) strain  is physically unacceptable in one-dimensional elastic or viscoelastic medium. Additionally, for large and positive values of $g(T)$ the small strain assumption of strain-limiting viscoelastic solid is violated. Therefore, we conclude that, in general, the quadratic and cubic models may result in either physically unacceptable  strain values or strain levels that are not consistent with the linearized strain assumption of strain-limiting theories. However, since they are the simplest representatives of the nonlinear models, for completeness we begin our discussion by considering general quadratic and cubic constitutive relations.

\subsection{Quadratic case}

For the quadratic case, we assume that the function $g(T)$ is of the form
\begin{equation}\label{quadratic}
    g(T) = g^{\prime}(0) T + \frac{1}{2} g^{\prime\prime}(0) T^{2}.
\end{equation}
We first consider the traveling wave  problem with the boundary conditions \eqref{bound} and then specify them to be as in \eqref{ass-pirs}. Substitution of \eqref{quadratic} into \eqref{c-2} gives
\begin{equation*}
    c^{2} = \left[g^{\prime}(0) + \frac{1}{2} g^{\prime\prime}(0)(T_{\infty}^{-} + T_{\infty}^{+})\right]^{-1}.
\end{equation*}
The restriction  $c^{2} > 0$ requires that   one of the following two cases must hold:
\begin{equation*}
    g^{\prime\prime}(0) > \frac{- 2 g^{\prime}(0)}{T_{\infty}^{-} + T_{\infty}^{+}} \qquad \text{and} \qquad T_{\infty}^{-}+T_{\infty}^{+} > 0,
\end{equation*}
or
\begin{equation*}
    g^{\prime\prime}(0) <  \frac{- 2 g^{\prime}(0)}{T_{\infty}^{-} + T_{\infty}^{+}} \qquad \text{and} \qquad T_{\infty}^{-}+T_{\infty}^{+} < 0.
\end{equation*}
With the use of \eqref{quadratic} in \eqref{general-ODE}, the differential equation we need to solve becomes the Riccati differential equation
\begin{equation} \label{riccati}
    T^{\prime} = a_{2} T^{2} + a_{1} T + a_{0},
\end{equation}
where $a_{0}$, $a_{1}$ and $a_{2}$ are constants defined by
\begin{equation*}
    a_{2} = - \frac{c g^{\prime\prime}(0)}{2 \nu}, \quad a_{1} = \frac{1 - c^{2} g^{\prime}(0)}{\nu c}, \quad a_{0} = -{1\over 2} \left(T_{\infty}^{+}+T_{\infty}^{-}\right) (1 - \theta)
\end{equation*}
with
\begin{equation*}
    \theta =  \frac{g^{\prime}(0) + \frac{1}{2} g^{\prime\prime}(0) \frac{(T_{\infty}^{-})^{2} + (T_{\infty}^{+})^{2}}{T_{\infty}^{+}+T_{\infty}^{-}}}{g^{\prime}(0) + \frac{1}{2} g^{\prime\prime}(0) (T_{\infty}^{-}+T_{\infty}^{+})}.
\end{equation*}
We observe that, when $T_{\infty}^{-}=0$ or $T_{\infty}^{+}=0$, the coefficient $a_{0}$ vanishes and \eqref{riccati} reduces to the Bernoulli differential equation. This makes it possible to find explicit solutions.

\begin{figure}[ht]
\centering
\begin{subfigure}{.5\textwidth}
  \centering
  \includegraphics[width=200pt]{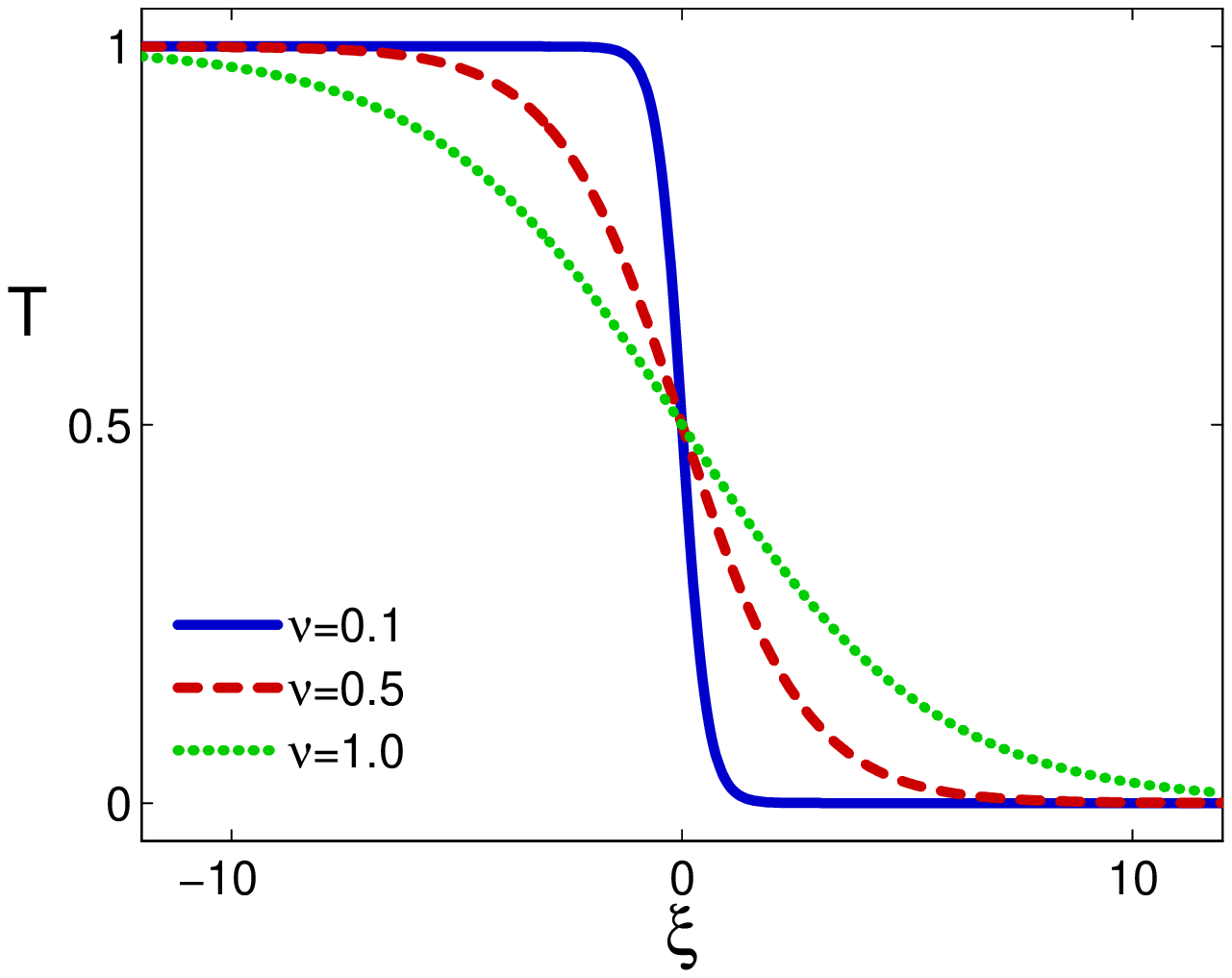}
  \caption{Stress}
  \label{fig:quadA}
\end{subfigure}%
\begin{subfigure}{.5\textwidth}
  \centering
  \includegraphics[width=200pt]{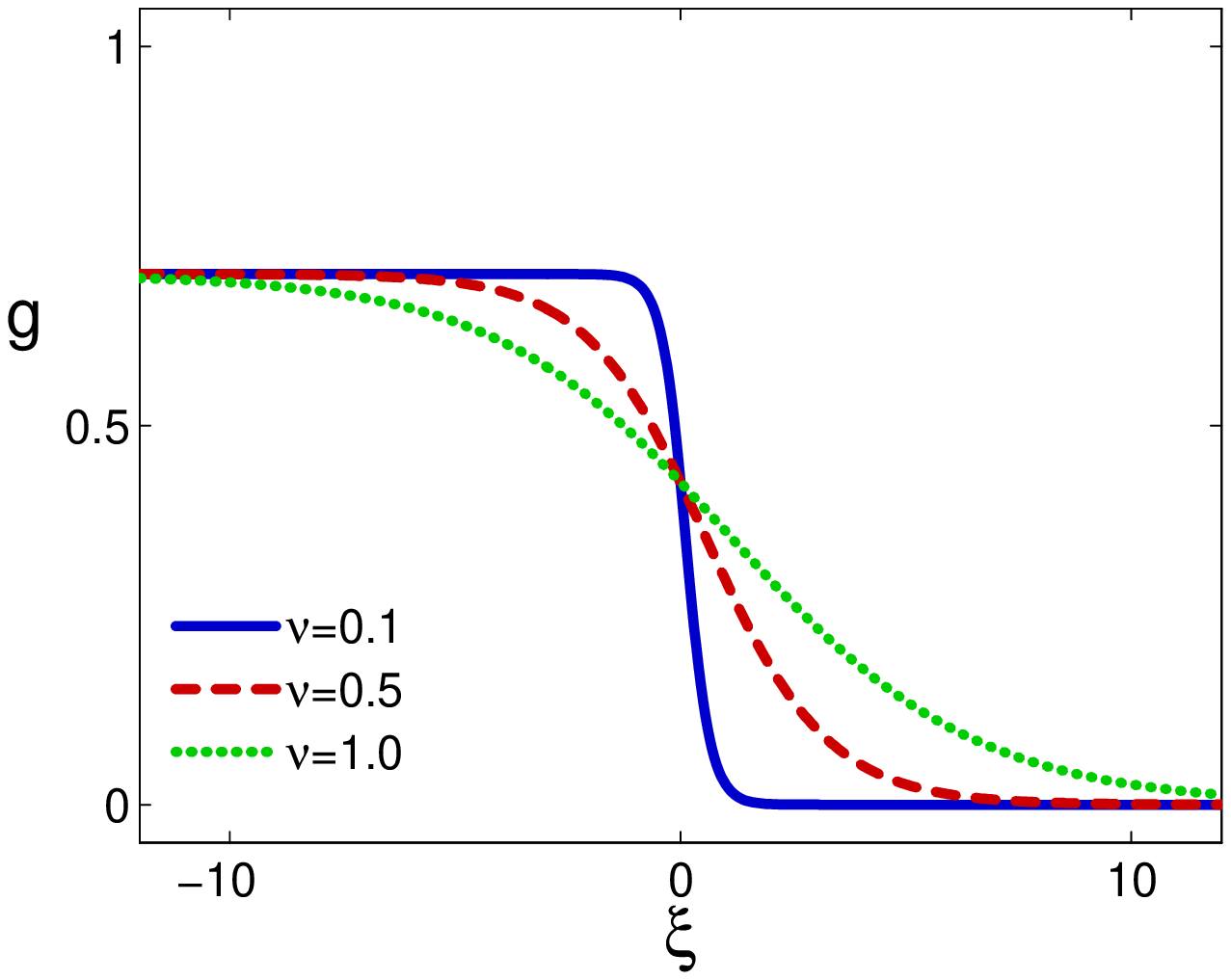}
  \caption{Strain}
  \label{fig:quadB}
\end{subfigure}
\caption{Variation of (a) $T(\xi)$, and (b) $g(T(\xi))$ of the quadratically nonlinear model  with $\xi$ for three different values of $\nu$ (where $g^{\prime}(0)=1$ and $g^{\prime\prime}(0)=-0.6$).}
\label{fig:quad}
\end{figure}

We now turn to the case of \eqref{ass-pirs} in which the wave propagation speed and the constants  $a_{0}$, $a_{1}$ and $a_{2}$  reduce to
\begin{equation*}
    c^{2} = \left[g^{\prime}(0) + \frac{1}{2}g^{\prime\prime}(0)  \right]^{-1} >0,
\end{equation*}
and
\begin{equation}\label{aaa}
    a_{2} = - \frac{g^{\prime\prime}(0) c}{2 \nu}, \quad a_{1} = - a_{2}, \quad a_{0} = 0,
\end{equation}
respectively. Consequently, \eqref{riccati} takes the form
\begin{equation}\label{bern}
    T^{\prime} = a_{2} T(1 - T),
\end{equation}
which admits the only two equilibrium solutions $T=0$ and $T=1$. Using \eqref{implicit} we find that, under the condition  \eqref{T_0}, the explicit solution of \eqref{bern} is found as
\begin{equation}\label{T}
    T(\xi) = \Big(1 + \exp (a_{2} \xi)\Big)^{-1}.
\end{equation}
The important point to note here is that \eqref{T} satisfies the conditions defined by \eqref{bound} and \eqref{ass-pirs} if  $a_{2}  > 0$. Combining this with \eqref{aaa}  implies that the traveling wave solution exists if  $g^{\prime\prime}(0) < 0$ and $c > 0$ or if $g^{\prime\prime}(0) > 0$ and $c < 0$. In other words, the  heteroclinic wave solution is a right-going traveling wave if $g^{\prime\prime}(0) < 0$ and a left-going wave if  $g^{\prime\prime}(0) > 0$. On the other hand, there is no heteroclinic traveling wave solution   for the quadratic model if  $a_{2}  < 0$, that is, if $g^{\prime\prime}(0)$ and $c$ have the same sign. Figure \ref{fig:quad} shows the variation of the analytical solution given in \eqref{T} with $\xi$ for three different values (corresponding to small, moderate and large values) of the viscosity parameter $\nu$, as well as the profile for the sum of the linearized strain and the strain rate. We observe from Fig. \ref{fig:quad} that, as it is expected, the traveling wave profiles become smoother as the viscosity increases. We also deduce from Fig. \ref{fig:quadB} that the profile for $g(T)$ is strongly distorted, in fact its values are reduced, due to the nonlinear dependence. Of course, this distortion can be intensified by choosing the values of the parameter $g^{\prime}(0)$ appearing in the constitutive relation properly. Furthermore, by choosing this parameter sufficiently small one could stay in the regime where the linearized strain assumption is valid.

We close this part by examining  the possibility of a shock wave (a traveling discontinuity). Differentiating the explicit solution $T(\xi)$ given by \eqref{T}, we get
\begin{equation}\label{T-deriv}
    T^{\prime}(\xi) = - \frac{ a_{2} \exp(a_{2} \xi)}{\big(1 + \exp(a_{2} \xi)\big)^{2}}.
\end{equation}
The effective width of the traveling wave is defined as $d=(T_{\infty}^{-}-T_{\infty}^{+})/\max |T^{\prime}(\xi)|$. Using  \eqref{ass-pirs} we conclude from \eqref{T} that the effective  width of the heteroclinic  traveling wave is $d=4/|a_{2}|$ for the quadratic model. From \eqref{aaa}, it follows that  $d = 8 \nu/|g^{\prime\prime}(0) c|$. Since the width $d$ is  proportional to the viscosity parameter $\nu$, it is natural to expect that the  wave profile becomes smoother as $\nu$ increases.  Furthermore, since the denominator of $T^\prime(\xi)$ in \eqref{T-deriv} is never zero, we  conclude that a shock wave does not form.

\subsection{Cubic case}

In this case we assume that
\begin{equation}\label{cubic}
    g(T) = g^{\prime}(0) T + \frac{1}{2} g^{\prime\prime}(0) T^{2} + \frac{1}{6} g^{\prime\prime\prime}(0) T^{3}.
\end{equation}
 Substitution of \eqref{cubic} into \eqref{general-ODE} yields
\begin{equation}\label{cubODE}
    T^{\prime} = \frac{1}{\nu \,c} \left\{\left(T - \frac{T_{\infty}^{+} + T_{\infty}^{-}}{2}\right) - c^{2} \left[g^{\prime}(0) T + \frac{1}{2} g^{\prime\prime}(0) T^{2} + \frac{1}{6} g^{\prime\prime\prime}(0) T^{3} - \frac{g(T_{\infty}^{+}) + g(T_{\infty}^{-})}{2}\right]\right\}.
\end{equation}
We again consider  the case defined by \eqref{ass-pirs}. So it follows from \eqref{cc} that
\begin{equation*}
    c^{2}= \left[g^{\prime}(0)  + \frac{1}{2} g^{\prime\prime}(0)  + \frac{1}{6} g^{\prime\prime\prime}(0)\right]^{-1} >0.
\end{equation*}
Using \eqref{ass-pirs} in \eqref{cubODE} (or using \eqref{cubic} in \eqref{sample}) we obtain the differential equation
\begin{equation}\label{cubic-eqn}
    T^{\prime} = a T (1 - T) (T + b),
\end{equation}
where the constants $a$ and $b$ are given by
\begin{equation}\label{a-b}
    a = -\frac{c\,g^{\prime\prime\prime}(0)}{6 \nu}\qquad \text{and}\qquad    b = 1 + 3 \frac{g^{\prime\prime}(0)}{g^{\prime\prime\prime}(0)}.
\end{equation}
It is worth pointing out that \eqref{cubic-eqn} admits three equilibrium solutions; $T=0$, $T=1$ and $T=-b$. Solving the differential equation  \eqref{cubic-eqn}  with  \eqref{T_0} gives the  closed-form solution
\begin{equation*}
    \frac{T^{1 + b}}{(1 - T)^{b} (T + b)} = \frac{1}{1 + 2 b} \exp(b (1+b) a \xi).
\end{equation*}

\begin{figure}[h!]
\centering
\begin{subfigure}{.5\textwidth}
  \centering
  \includegraphics[width=200pt]{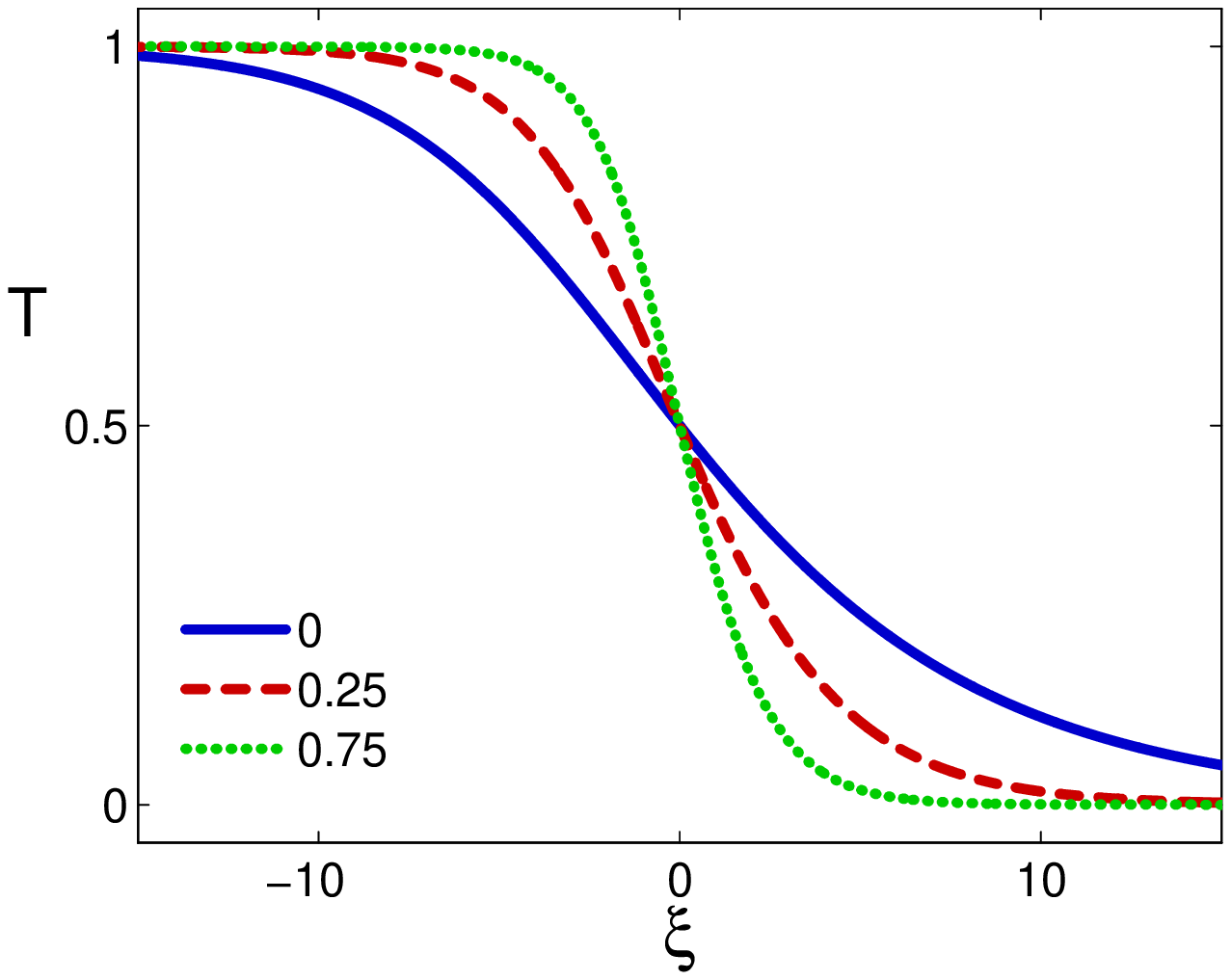}
  \caption{Stress}
  \label{fig:cubicA}
\end{subfigure}%
\begin{subfigure}{.5\textwidth}
  \centering
  \includegraphics[width=200pt]{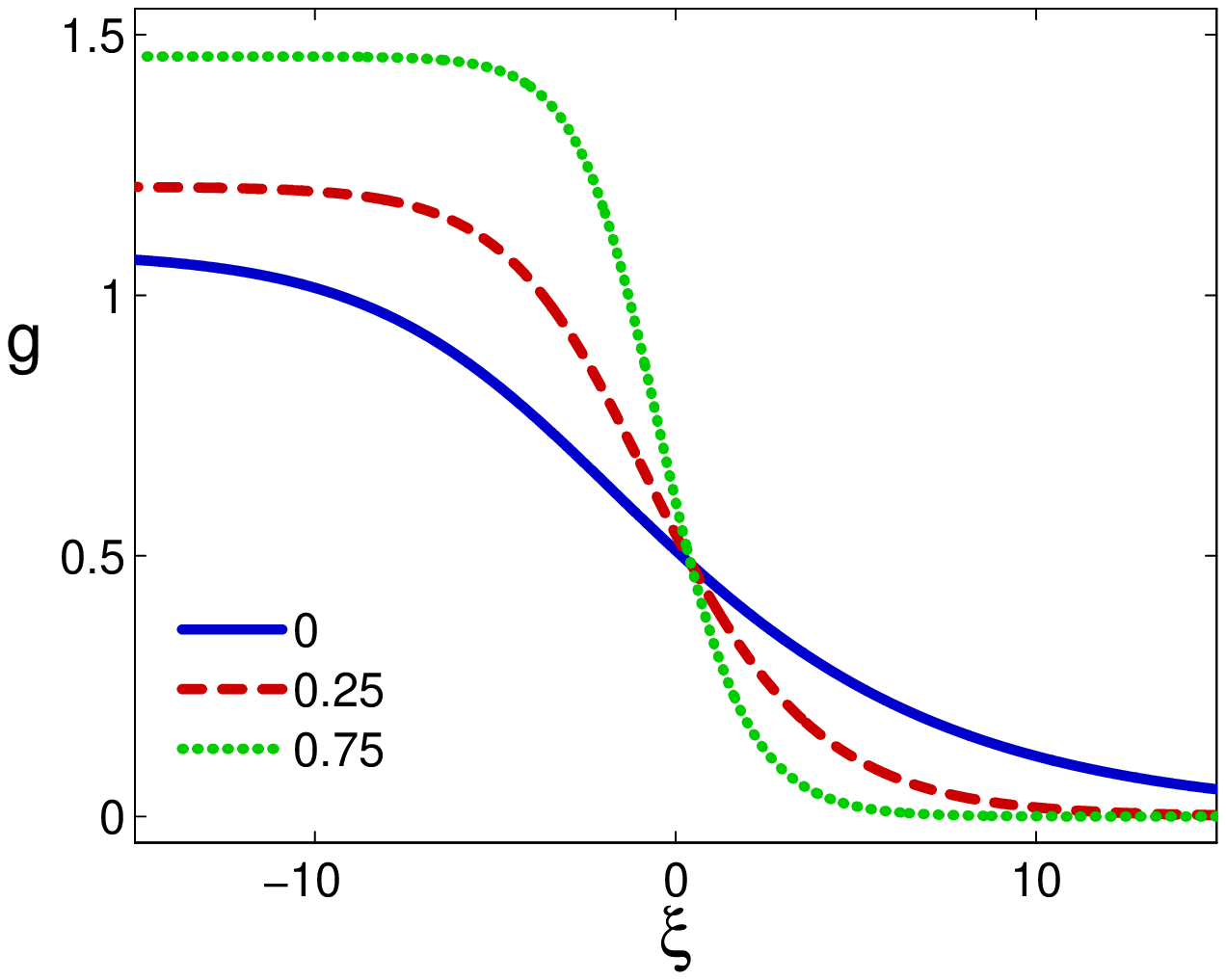}
  \caption{Strain}
  \label{fig:cubicB}
\end{subfigure}
\caption{Variation of (a) $T(\xi)$, and (b) $g(T(\xi))$ of the cubically  nonlinear model  with $\xi$ for different values of $g^{\prime\prime}(0)$  ($\nu=0.5$, $g^{\prime}(0)=1$ and $g^{\prime\prime\prime}(0)=0.5$). The curves on the graph correspond to three different values: $g^{\prime\prime}(0)=0, 0.25, 0.75$ (or equivalently $b=1, 2.5, 5.5$).}
\label{fig:cubic}
\end{figure}

Figure \ref{fig:cubic} presents the variation of $T(\xi)$ and $g(T(\xi))$ with $\xi$ for three different values of $g^{\prime\prime}(0)$, namely, $g^{\prime\prime}(0)=0, 0.25, 0.75,$ when $\nu=0.5$, $g(0)=1$ and $g^{\prime\prime\prime}(0)=0.5$. Note the different scales for the vertical axes of Fig. \ref{fig:cubicA} and Fig. \ref{fig:cubicB}. We observe that the profiles for the stress and the strain become smoother as $g^{\prime\prime}(0)$ (or equivalently $b$) increases. Similar to the quadratic case the profile for $g(T)$ is distorted due to the nonlinearity. Also, as it is expected from the behaviour of the cubic nonlinearity in Fig. \ref{lin-quad-cub}, the values of $T$ are amplified in Fig. \ref{fig:cubicB}. Furthermore, we note that, when $g^{\prime\prime}(0) = 0$ (or equivalently $b=1$), it is possible to obtain an explicit solution from the implicit one as
\begin{equation} \label{Tcube}
    T(\xi)= \frac{\exp(a \xi)}{(3 + \exp(2 a \xi) )^{1/2}}.
\end{equation}
The crucial fact is that the conditions given by \eqref{bound} impose a restriction on the constant $a$, which is the condition $a < 0$. This implies that, for the special cubic model, the traveling wave solution exists if $a < 0$, or equivalently, if $g^{\prime\prime\prime}(0)$ and $c$ have the same sign. The variation of the analytical solution given in \eqref{Tcube} with $\xi$ for three different values of the viscosity parameter $\nu$ produces a figure, which is very similar to Fig. 3 and shows that the same conclusions are also valid for the present case, and therefore we do not reproduce it here.

\subsection{Case of Model A}\label{sec:Kan-Raj-Sac}

We now take \eqref{modelA} to define the constitutive relation for our strain-limiting viscoelastic solid through \eqref{cons-eqn}. Of course, if $n=0$, \eqref{modelA} gives the linear model of strain-limiting viscoelasticity, for which we have already mentioned that there is no heteroclinic traveling wave. In general, depending on the values of the parameters appearing in \eqref{modelA} the function $g(T)$ exhibits very different patterns of behaviour.  As stated before, when $\beta = 0$, $n=-1/2$, $\alpha=1$ and $\gamma=2$, Model A becomes equivalent to Model B with $r=2$ and the implicit solution corresponding to this special case is given in the next subsection. In this subsection we restrict our attention to the case $n=1$, which allows us to find an explicit solution for the corresponding differential equation.  Substituting  \eqref{modelA} with $n=1$ into \eqref{sample} yields
\begin{equation}\label{K-ode}
    T^{\prime} = \kappa T (1 - T^{2}),
\end{equation}
where
\begin{equation}
    \kappa = \frac{\alpha\gamma}{[\alpha(1 + \gamma)+\beta] \nu c}.
\end{equation}
It is clear that \eqref{K-ode} is a special case of \eqref{cubic-eqn}, with $b = 1$ and $ a = \kappa$. Therefore, if we replace $a$ in \eqref{Tcube} by $\kappa$ we get the explicit solution corresponding to \eqref{modelA} with $n=1$. Additionally, we conclude that  the traveling wave solution exists if $\kappa<0$, or equivalently if $c [\alpha(1 + \gamma)+\beta] < 0$ (recall that $\nu >0, \alpha \geq 0, \gamma \geq 0$). Since the traveling wave solution is a special case of \eqref{Tcube}, we can draw the same conclusions by saying that the traveling wave profiles become smoother as the viscosity increases as well as the wave profile for $g(T)$ is distorted due to the nonlinearity.

\subsection{Case of Model B}\label{sec:modelB}

\begin{figure}[h!]
\centering
\begin{subfigure}{.5\textwidth}
  \centering
  \includegraphics[width=200pt]{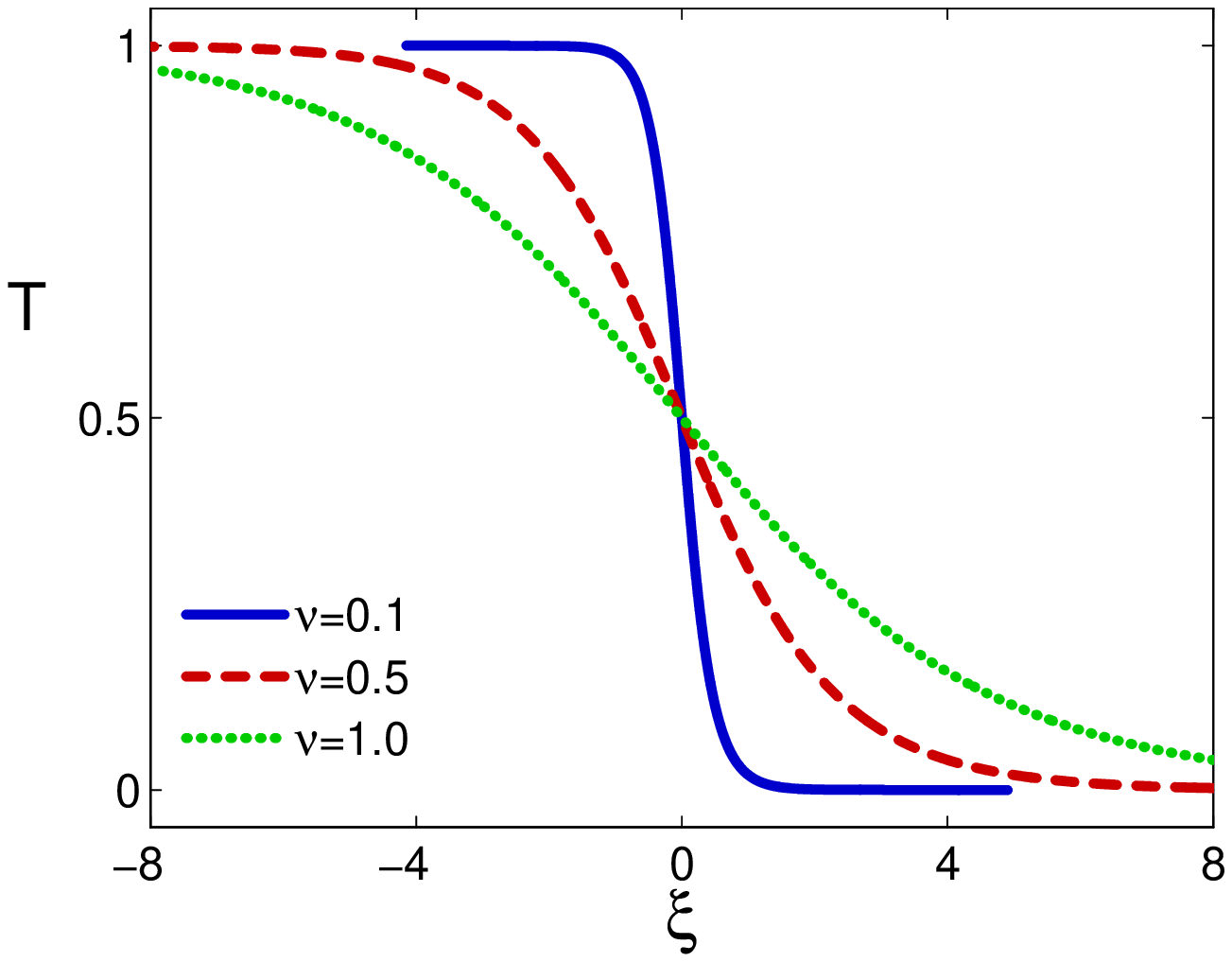}
  \caption{Stress}
  \label{fig:modelBa}
\end{subfigure}%
\begin{subfigure}{.5\textwidth}
  \centering
  \includegraphics[width=200pt]{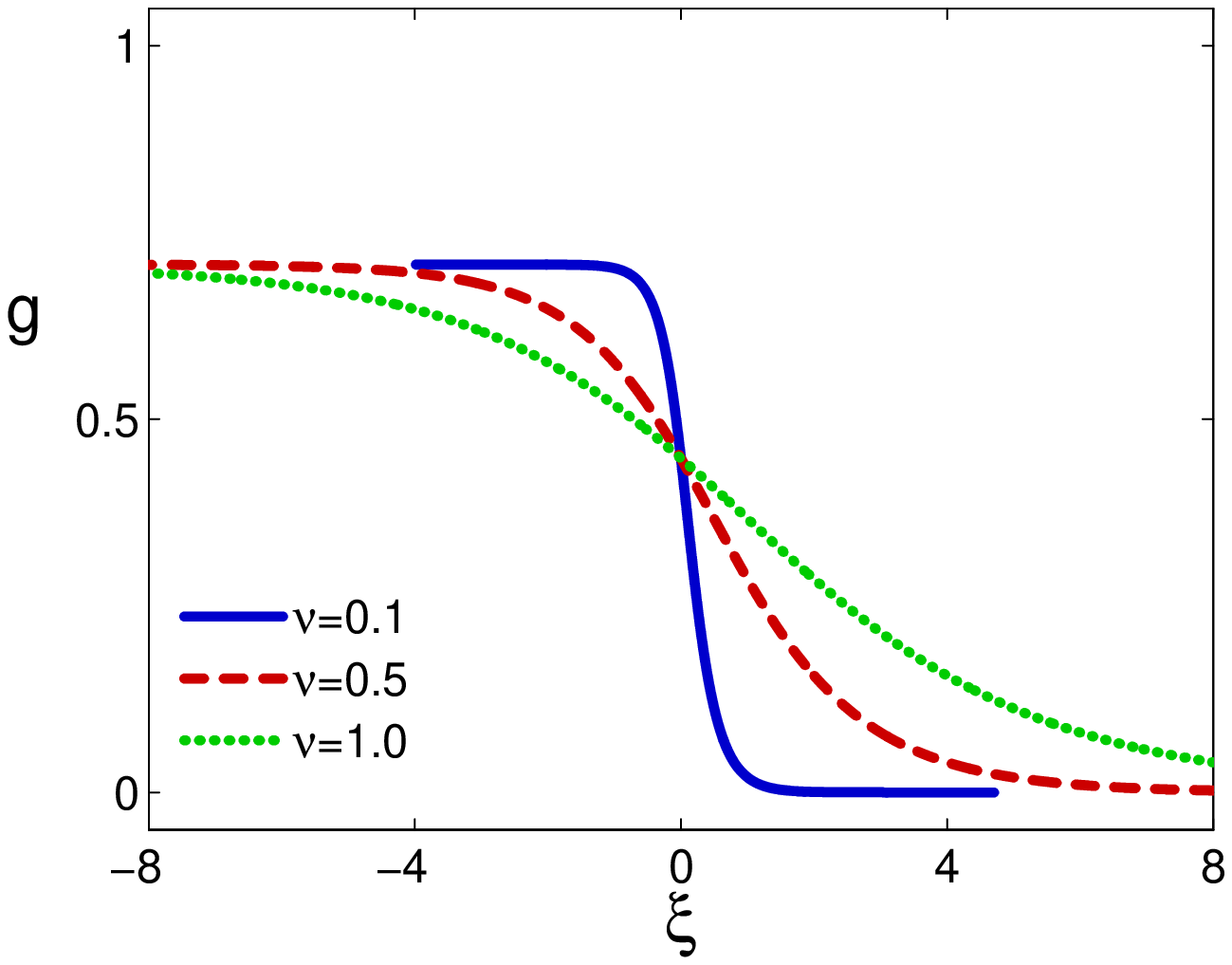}
  \caption{Strain}
  \label{fig:modelBb}
\end{subfigure}
\caption{Variation of (a) $T(\xi)$, and (b) $g(T(\xi))$ of Model B with $\xi$ for three different values of $\nu$ ($r=2$).}
\label{fig:modelB}
\end{figure}

Here we take \eqref{modelB} as $g(T)$ in \eqref{cons-eqn}.  Note that combining \eqref{modelB} with \eqref{cc} gives $c^{2}=2^{1/r}$. If we substitute \eqref{modelB} into \eqref{sample}, we get the differential equation that we need to solve as
\begin{equation}\label{Suli-eqn}
    T^{\prime} = \frac{T}{\nu c} \left(1 - \frac{2^{1/r}}{(1 + | T |^{r})^{1/r}} \right).
\end{equation}
In the case of $r = 2$, we find an analytical solution of \eqref{Suli-eqn}. Using \eqref{implicit} and \eqref{T_0} we obtain the solution implicitly as
\begin{equation}\label{Hfunc}
    H(T) = H(1/2) \exp(\xi/\nu c) ,
\end{equation}
where the function $H(s)$ is defined as
\begin{equation*}
    H(s) = \left(\frac{(1 - s^{2})^{2}}{s [3 + s^{2} + 2^{3/2} (1 + s^{2})]} \right) \left(\frac{(1+s^{2})^{1/2} +1}{s}\right)^{2^{1/2}}.
\end{equation*}
Note that some basic properties of $H(s)$ are as follows:
\begin{equation*}
    H(\mp 1) = 0, \qquad \quad H(1/2) > 0,
\end{equation*}
and
 \begin{equation*}
    H(s) \to \infty ~~\mbox{as}~~ s\to 0^{+}.
\end{equation*}
By combining these properties and  \eqref{Hfunc}, we deduce  the following two sets of results, depending on the sign of $c$. If $c>0$, we get
\begin{equation*}
  T \to 0^{+} \quad\text{as}\quad \xi \to + \infty  ~~~\mbox{and}~~~ T \to \mp 1 \quad\text{as}\quad \xi \to - \infty.
 \end{equation*}
Similarly, if $c<0$, we get
 \begin{equation*}
 T \to \mp 1 \quad\text{as}\quad  \xi \to + \infty  ~~~\mbox{and}~~~ T \to 0^{+} \quad\text{as}\quad \xi \to - \infty.
 \end{equation*}
We restrict our attention to the case $c > 0$ since the corresponding conditions are compatible with \eqref{bound} and \eqref{ass-pirs}.

Figure \ref{fig:modelB} shows the graph of the implicit solution for a specific set of parameter values and for three different values of the viscosity parameter $\nu$, and also the profile of the strain. From Fig. \ref{fig:modelB}, we can see that the traveling wave profile becomes smoother as the viscosity increases, and the profile for $g(T)$ is distorted due to the nonlinearity as in the previous models.

\begin{figure}[h!]
\centering
\begin{subfigure}{.5\textwidth}
  \centering
  \includegraphics[width=200pt]{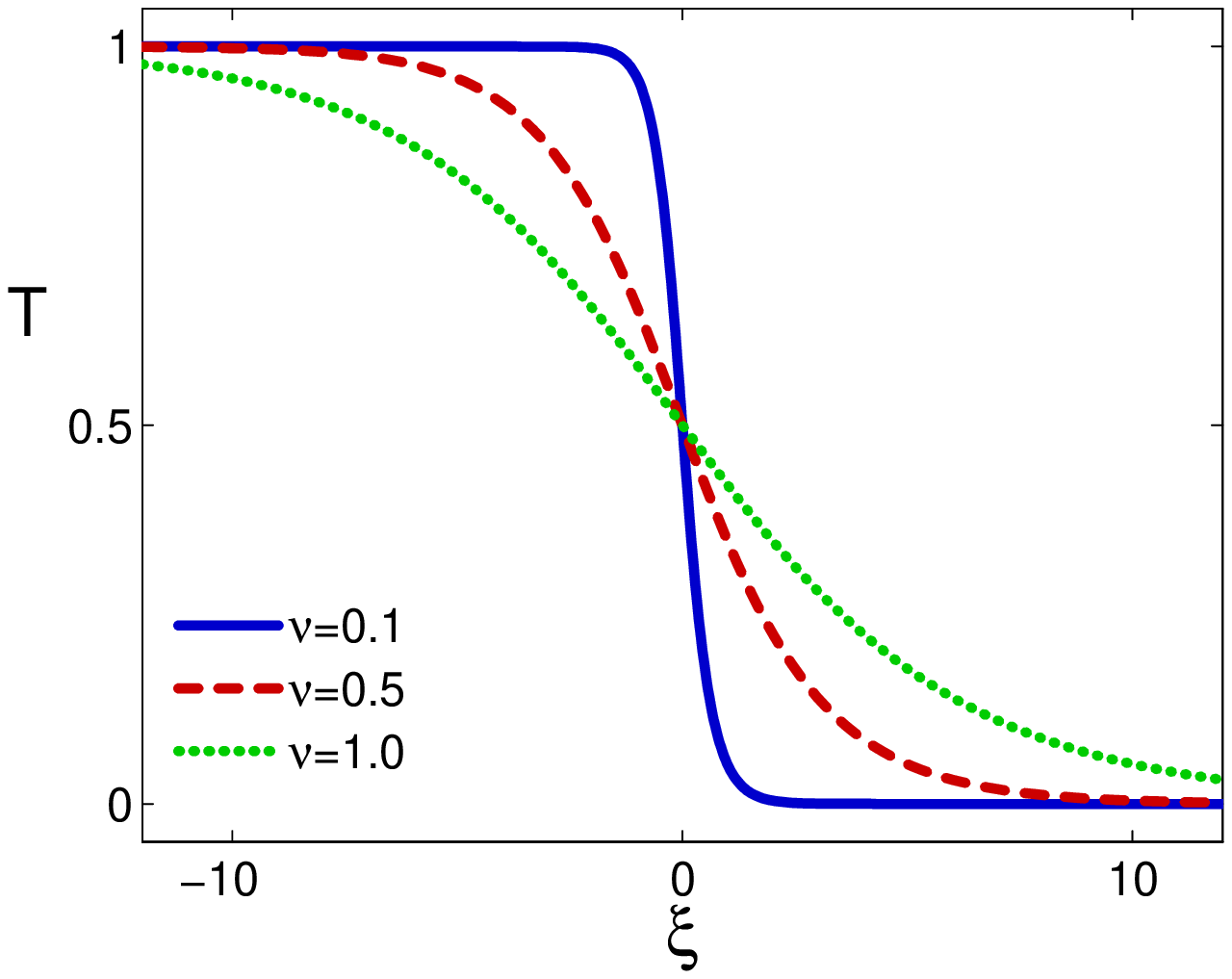}
  \caption{Stress}
  \label{fig:modelCa}
\end{subfigure}%
\begin{subfigure}{.5\textwidth}
  \centering
  \includegraphics[width=200pt]{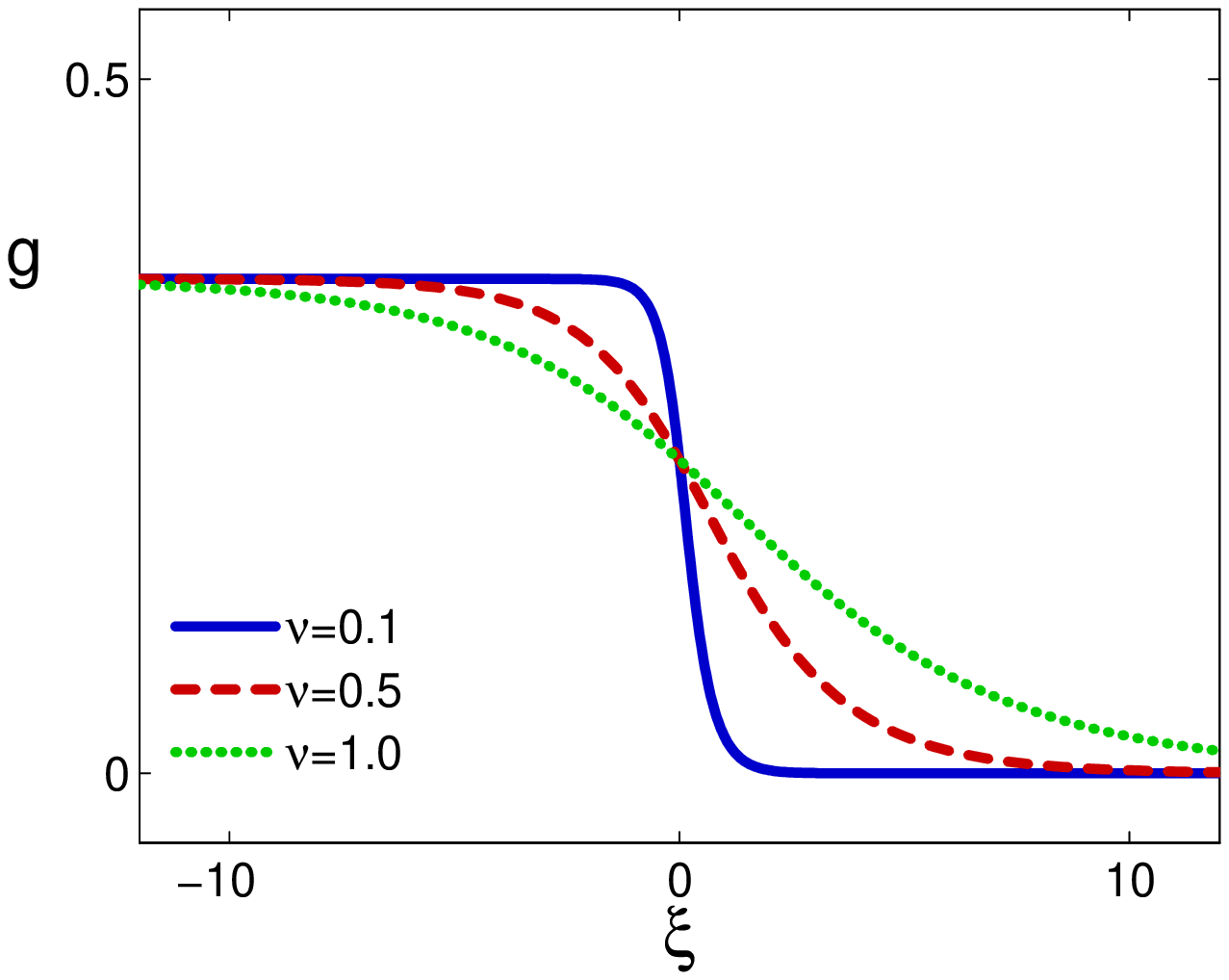}
  \caption{Strain}
  \label{fig:modelCb}
\end{subfigure}
\caption{Variation of (a) $T(\xi)$, and (b) $g(T(\xi))$ of Model C  with $\xi$ for three different values of $\nu$  ($\alpha=0.5$, $\beta=0.01$ and $\gamma=\delta=1$).}
\label{fig:modelC}
\end{figure}

\subsection{Case of Model C}\label{sec:modelC}

We obtain the differential equation we need to solve by substituting  \eqref{modelC} into \eqref{sample}. However, since the resulting equation is highly nonlinear and an analytical solution is not available, we focus on the numerical solution instead. To this end we use MATLAB function \texttt{ode45} to solve the differential equation, which is the standard solver of MATLAB for ordinary differential equations. Omitting the details of the numerical calculations, we show in Fig. \ref{fig:modelC} the numerical solutions for three different values of the viscosity parameter $\nu$.

\subsection{Case of Model D}\label{sec:modelD}

\begin{figure}[h!]
\centering
\begin{subfigure}{.5\textwidth}
  \centering
  \includegraphics[width=200pt]{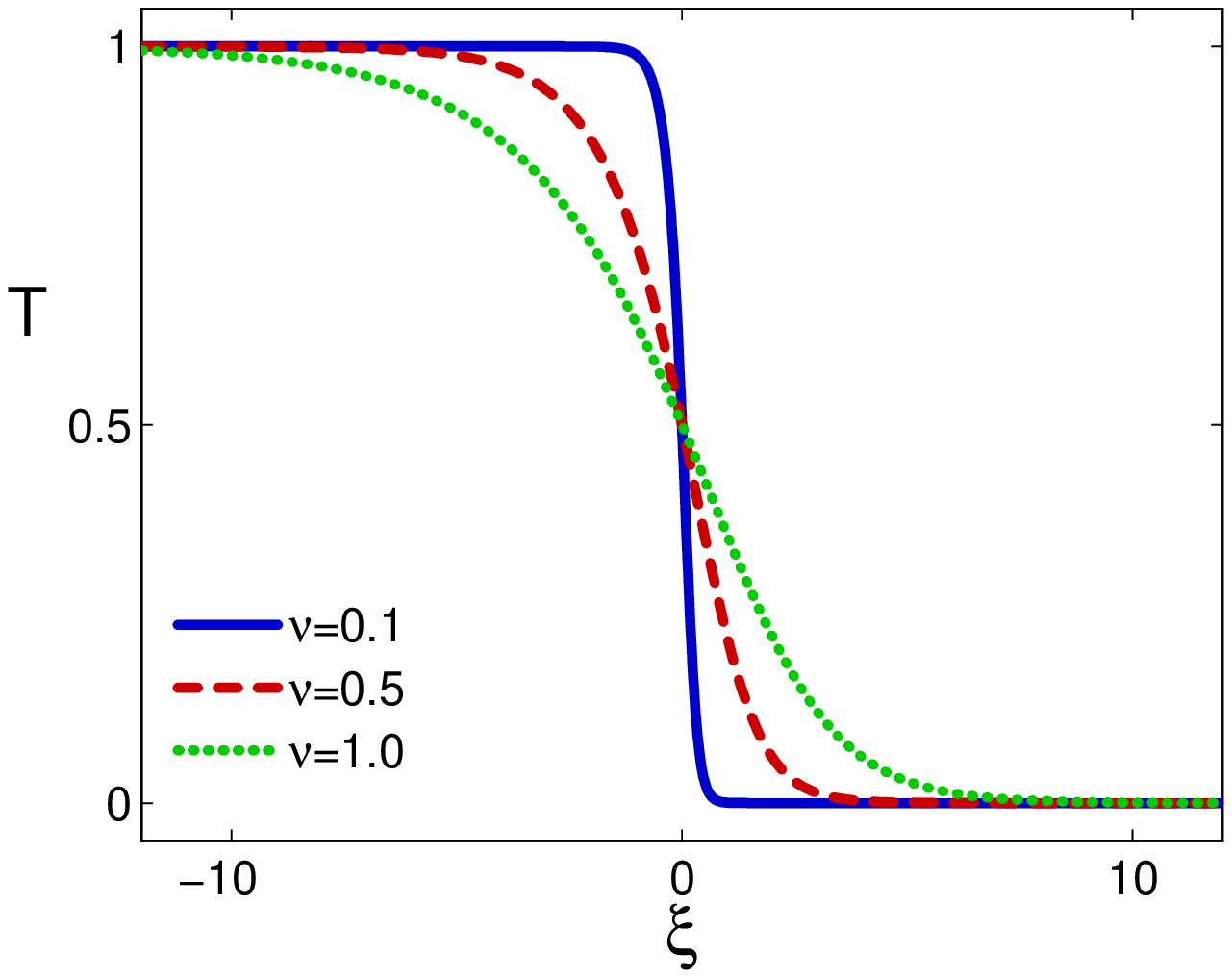}
  \caption{Stress}
  \label{fig:modelDa}
\end{subfigure}%
\begin{subfigure}{.5\textwidth}
  \centering
  \includegraphics[width=200pt]{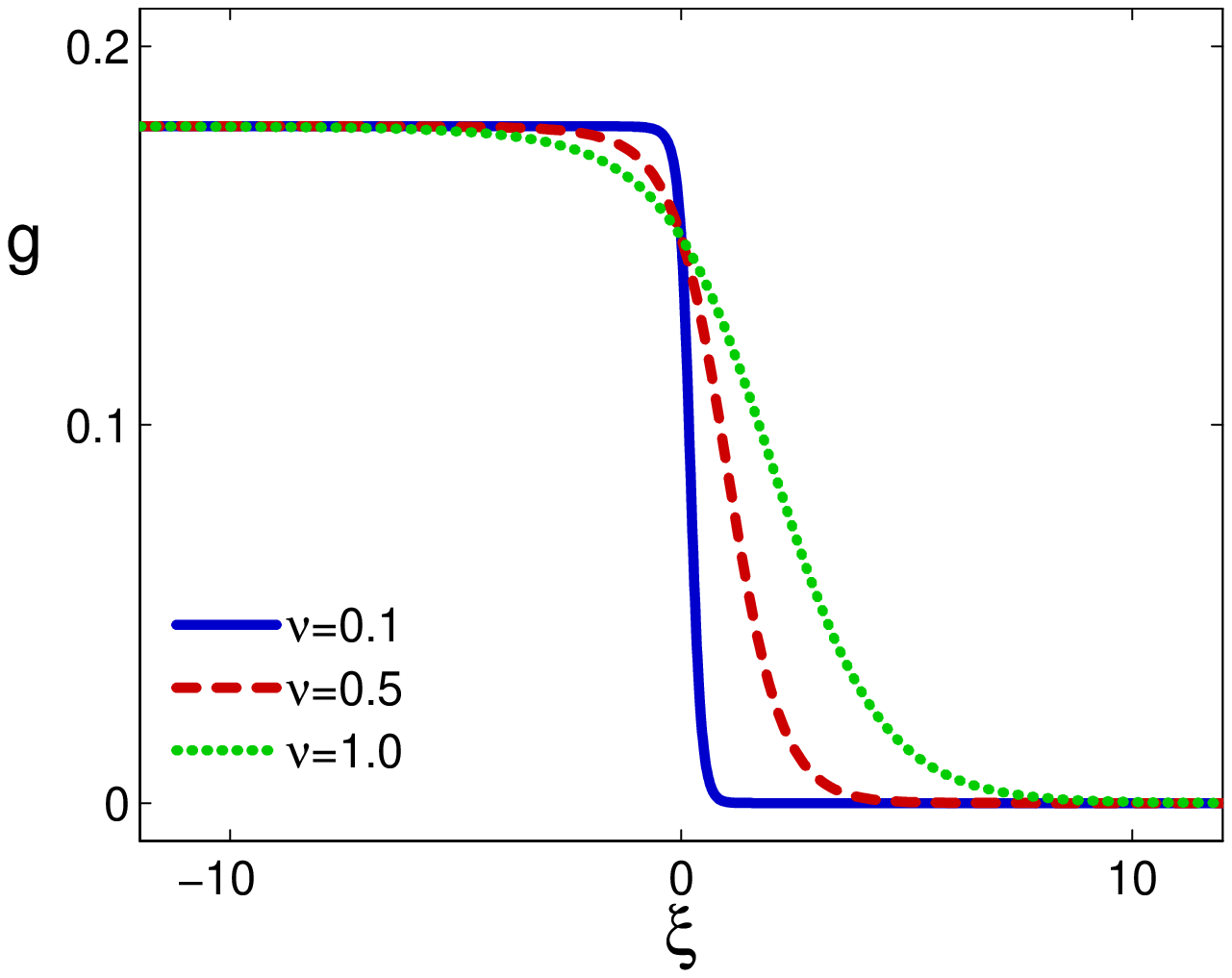}
  \caption{Strain}
  \label{fig:modelDb}
\end{subfigure}
\caption{Variation of (a) $T(\xi)$, and (b) $g(T(\xi))$ of Model D  with $\xi$ for three different values of $\nu$ ($\alpha=0.5$, $\beta=0.01$, $\gamma=\delta=1$ and $n=0.5$) .}
\label{fig:modelD}
\end{figure}

When we substitute  \eqref{modelD} into \eqref{sample} we again get a highly nonlinear differential equation for which an analytical solution is not possible to find. For this reason, it is convenient to solve it numerically using MATLAB function \texttt{ode45} just as above. In Fig. \ref{fig:modelD} we plot the numerical solutions for three different values of the viscosity parameter $\nu$.

Both Fig. \ref{fig:modelC} and Fig. \ref{fig:modelD} clearly show that Models C and D have kink-type traveling wave solutions and that the wave profiles obtained numerically for the stress are in qualitatively good agreement with those derived from the analytical solutions belonging to the previous models. We note that the wave profiles for the strain are significantly different from those of the previous models. The remarks made for those models regarding the smoothness and distortion of the profiles are also valid in both cases. We also observe that the reduction of the values of $g(T)$ in Model D is significantly stronger than that of Model C.

\newpage
\bibliographystyle{plainnat}
\bibliography{bibliography}

\end{document}